\documentclass[11pt]{article}
\usepackage{a4}
\usepackage{epsf}
\begin{document}
\title{How to compute Green's Functions for entire\\
       Mass Trajectories within Krylov Solvers}
\author{U. Gl\"assner$^{a}$, S. G\"usken$^{a}$, Th. Lippert$^{b}$,\\ 
        G. Ritzenh\"ofer$^{b}$, K. Schilling$^{a,b}$, 
        and A. Frommer$^{c}$\\[0.4cm]
        \footnotesize
        $^{a}$Department of Physics, University of Wuppertal,
        42097 Wuppertal, Germany\\
        \footnotesize
        $^{b}$HLRZ, c/o KFA-J\"ulich, D-52425 J\"ulich,
        Germany and DESY, Hamburg, Germany\\
        \footnotesize
        $^{c}$Department of Mathematics, University of Wuppertal,
        42097 Wuppertal, Germany}
\maketitle
%Glaessner       Feb    1st version of the algorithm
%Glaessner       May 2, algorithm  
%Lippert         May 3, intro, abstract
%Glaessner       May 4, algorithm and measurements
%Glaessner       May 5, added plots
%Glaessner       May 6, minor text modifications
%Schilling, Lippert, Glaessner  May 6, larger ;-) text modifications
%Glaessner       May 7, added eo-part
%Guesken,Glaessner May 7, one-before-last corrections
% May 8 final version ?????????

\begin{abstract}
  The availability of efficient Krylov subspace solvers play a vital
  role for the solution of a variety of numerical problems in
  computational science. Here we consider lattice field theory.
  
  We present a new general numerical method to compute many
  Green's functions for complex non-singular matrices within one
  iteration process.  Our procedure applies to matrices of structure
  $A=D-m$, with $m$ proportional to the unit matrix, and can be
  integrated within any Krylov subspace solver.  We can compute the
  derivatives $x^{(n)}$ of the solution vector $x$ with respect to the
  parameter $m$ and construct the Taylor expansion of $x$ around $m$.
  
  We demonstrate the advantages of our method using a minimal residual
  solver. Here the procedure requires $1$ intermediate vector for each
  Green's function to compute.  As real life example, we determine a
  mass trajectory of the Wilson fermion matrix for lattice QCD. Here
  we find that we can obtain Green's functions at all masses $\geq m$
  at the price of one inversion at mass $m$.
\end{abstract}

\section{Introduction}
Lattice Quantum Chromodynamics demands for enormous compute power and
efficient numerical algorithms in order to solve huge sparse systems
of linear equations. These linear systems originate from a discretized
differential operator, the fermionic matrix, with varying gluonic
background fields acting as coefficient functions. The latter can be
generated by Monte Carlo methods. From the solutions, i.e. the Green's
functions or propagators, one can determine the properties of hadrons,
as e.g.\ spectrum, weak decay constants or weak matrix elements
\cite{MONTVAY,Cr83,Ro92}.

Recently, in Ref.~\cite{FROMMER95}, it has been demonstrated that it
can be very advantageous to exploit the structure of the Wilson
fermion matrix $M=m-D$ \cite{WILSON}, which is widely used in lattice
calculations.  In particular, the computation of so-called mass
trajectories, i.e., multiple inversions of $m-D$ for a set of `bare
masses'\footnote{The 'physical' bare masses differ from $m$ by a
  constant shift, which is not relevant in this context} $m_0, m_1,
\ldots, m_N$ with $m_i < m_{i+1} $, could be carried out within one
iteration process, using the quasi minimum residual algorithm (QMR)
\cite{FN91,FREUND93-2,FROMMER94}.  This useful property is a direct
consequence of the non-symmetric Lanczos process \cite{LANCZOS50}, where
the Lanczos part had to be gone through only once as it depends solely
on $D$.  Therefore, the method presented in Ref.~\cite{FROMMER95} is
restricted to QMR. Unfortunately the application of the QMR-MULT
algorithm \cite{FROMMER95} is rather memory intensive: QMR itself
needs $6$ intermediate vectors of the size of one propagator component
and requires to store $3$ additional vectors for each of the members
of the mass-trajectory.  Computations of propagators as of today,
however, tend to be memory bounded even on the largest distributed
memory machines available.

In this paper, we propose a {\em new} numerical approach to compute
the entire mass trajectory simultaneously.  This method is not
restricted to a special solver such as QMR and requires only a modest
additional amount of workspace.

For definiteness and simplicity we present our idea as applied within
the minimal residual algorithm (MR), one of the favorite standard
solvers in lattice QCD that belongs to the class of the general
conjugate residual methods\cite{EISEN, Oy85}. MR is a GCR algorithm with
recursion depth $0$.

Our key observation is that the derivatives $x^{(n)}=d^nx/dm^n$ of the
solution $x$ can be related recursively to the coefficients that occur
within the iteration. Thus the Taylor expansion of the propagator
$x(m+\Delta m)$ around $m$ can be computed almost for free during the
iteration process.

We find numerically that the procedure renders $x(m_i>m_0)$
automatically with higher accuracy than $x(m_0)$. The accuracy of the
latter is determined as usual by the stopping residual's norm.  This
enables us to calculate the entire mass trajectory at once, provided
that we compute at $m_0$. On the other side, the procedure can provide
good estimates for propagators with $x(m_i<m_0)$ to be used as
educated guesses for further inversions.

We note that our method can be applied to the case of
even-odd-preconditioning \cite{DEGRAND}. For general sources, however,
two separate inversions are necesssary. The additional expense of a
factor of two easily pays off against large numbers of Green's
functions to be computed.

As a by-product we are in principle in the position to compute all
derivatives $x^{(n)}$ up to order $n=N$, the number where the
iteration stops.

The paper is organized as follows: In section \ref{DER:ALG}, we
present the MR algorithm followed by the computation of the
derivatives in section \ref{DER:DER}. The construction of the Taylor
series is explained in section \ref{DER:TAY}.  Furthermore we will
present the results of numerical tests that deal with the inversion of
the Wilson fermion matrix on a real full QCD configuration at
$\beta=5.6$ and $\kappa=0.1575$ on a $16^{3}\times 32$ lattice.

\section{M$^{3}$R: Multiple Masses Minimal Residual}

\subsection{Standard Minimal Residual Algorithm}
\label{DER:ALG}
Consider the linear system: 
\begin{equation}
A x = (D-m) x = r_0 \label{equation}
\end{equation}
with $A$ being a complex nonsingular matrix.  The search direction in
the Minimal Residual Algorithm is taken to be the direction of the
residual. At every iteration step the following operation is performed:
\begin{eqnarray}
x_{i+1} & = & x_i + \alpha_i r_i \\
r_{i+1} & = & r_i - \alpha_i A r_i . \nonumber
\end{eqnarray}
The coefficient $\alpha_i$ is determined by a minimization of the
2-norm of the new residual $r_{i+1}$:
\begin{equation}
\alpha_i = \frac{(A r_i)^{\dagger} r_i}{(A r_i)^{\dagger} A r_i} .
\end{equation}

\subsection{Computation of the Derivatives}
\label{DER:DER}
It is possible to extract in addition to the solution an arbitrary
number of derivatives $x^{(n)}=d^n x(m)/dm^n$ at almost no extra
computational cost, provided the matrix is of the special form of
equation \ref{equation} and the source $r_0$ is independent of $m$.

$n$-fold differentiation of eq. \ref{equation} with respect to $m$
leads to a linear equation for the $n$-th derivative
\begin{equation}
A x^{(n)} - n x^{(n-1)} = 0 . \label{der_equation} 
\end{equation}
Let $x_i$ denote the approximation of the solution $x$ after $i$
iterations of the Minimal Residual algorithm. We are interested in a
similar approximation for the 1st derivative $x_i^{(1)}$. Therefore we
demand eq. \ref{der_equation} for $n=1$ to hold at each iteration
step:
\begin{equation}
A x_i^{(1)} = x_i  
\end{equation}
It is obvious that the approximants $x_i^{(1)}$, defined in this way,
converge towards $x^{(1)}$ as $x_i$ approaches $x$.  The next step
is to calculate the recursion relation for $x_i^{(1)}$.  To this end
we make use of the recursion relation for $x_i$ itself which is
particularly simple in the case of the Minimal Residual algorithm:
\begin{eqnarray}
A (x_{i+1}^{(1)} - x_i^{(1)}) 
= (x_{i+1} - x_{i}) & = & \alpha_i r_i \nonumber\\
                    & = & \alpha_i (r_0 - A x_i) . 
\end{eqnarray}
We then find the formal recursion relation 
\begin{equation}
 x_{i+1}^{(1)} - x_{i}^{(1)}  = \alpha_i (x - x_i)  . \nonumber 
\end{equation}
At this stage we replace the true solution $x$ by it's best 
approximation $x_N$.
\begin{equation} 
  x_{i+1}^{(1)} - x_{i}^{(1)}  = \alpha_i \left[ \sum_{j=0}^N 
\alpha_j r_j - \sum_{j=0}^{i-1} \alpha_j r_j \right] 
\end{equation}
A summation over $i$ then leads to the following expression for 
$x_{N+1}^{(1)}$:
\begin{equation}
x_{N+1}^{(1)} - x_0^{(1)} = \sum_{i=0}^N \alpha_i \sum_{j=i}^N
\alpha_j r_j .
\end{equation}
It is important to start the inversion with an initial guess $x_0=0$
since the nonsingularity of $A$ then implies that $x_0^{(1)}$
vanishes as well\footnote{A nontrivial starting guess for $x_0$ would
  require a starting guess for $x_0^{(1)}$ computed from $A x_0^{(1)} =
  x_0$ at the price of an extra inversion}. After a resummation
of the r.h.s we find
\begin{equation}
x_{N+1}^{(1)} = \sum_{i=0}^N \alpha_i \left[ \sum_{j=0}^i \alpha_j
\right]  r_i .
\end{equation}
Note that we can recursively build up the r.h.s. of this equation 
in terms of the coefficients
\begin{equation}
\alpha_i^{(1)} = \alpha_i \sum_{j=0}^i \alpha_j
\end{equation}
at the price of only one extra vector-update per iteration step
\begin{equation}
x^{(1)}_{i+1} =  x^{(1)}_i + \alpha_i^{(1)} r_i .
\end{equation}

One can extend this procedure to higher derivatives by use of the
following recursion relation for the generalized coefficients:
\begin{equation}
\alpha_i^{(n+1)} = \alpha_i \sum_{j=0}^i \alpha_j^{(n)} .\label{alpha_def}
\end{equation}
In this way one ends up with the general updating-step for the derivatives
\begin{equation}
x^{(n)}_{i+1} = x^{(n)}_i + n! \alpha_i^{(n)} r_i . \label{derivatives}
\end{equation}

We remark that for every additional derivative only one extra vector 
of storage is necessary.

\subsection{Recombination of the Taylor-series}
\label{DER:TAY}
Usually one is interested in the solution of eq. \ref{equation} for a
given set of different values of $m$. In this section we will show
that it is possible to perform a summation over all available
approximate derivatives of $x(m)$ and hence obtain a Taylor
extrapolation to a different value of the parameter $m$.

We would like to perform a summation of the N-truncated Taylor series
to an extrapolated solution $x^E \equiv x(m + \Delta m)$:
\begin{equation}
x^E = \sum_{n=0}^N \frac{(\Delta m)^n}{n!} x^{(n)} .
\end{equation}
This Taylor series can be constructed recursively by use of
equation \ref{derivatives}:
\begin{equation}
x_{i+1}^E  = x_i^E + \sum_{n=0}^N (\Delta m)^n \alpha_i^{(n)} r_i
          := x_i^E + f_i \alpha_i r_i , 
\end{equation}
where we have introduced scale-factors 
\begin{equation}
f_i = 1 + \frac{1}{\alpha_i} \sum_{n=1}^N \Delta m^n \alpha_i^{(n)} .
\label{update}
\end{equation}

In order to compute these scale-factors $f_i$ we recast eq.
\ref{alpha_def} in recursive form:
\begin{equation}
\alpha_i^{(n)} = \frac{\alpha_i}{\alpha_{i-1}}\alpha_{i-1}^{(n)} 
                 + \alpha_i \alpha_{i}^{(n-1)} .
\end{equation}
Inserted into equation \ref{update} this yields:
\begin{eqnarray}
f_i & = & \frac{f_{i-1}}{1-\Delta m \alpha_i} \\
f_0 & = & \frac{1}{1-\Delta m \alpha_0} \nonumber .
\end{eqnarray}
We close with the comment that the entire procedure is selfconsistent
up to terms of $O(m^{N+1})$ and $O(\Delta m^{N+1})$.

\subsection{Template for the $M^3R$-Algorithm}
Next we write down a template for our Multiple Masses Minimal Residual
$(M^3R)$ algorithm.  For shortness we apply one extrapolated solution
$x^E(\Delta m)$, more solutions can be included easily.

\hspace{1cm}

\begin{center}
\begin{tabular}{ll}
Initialize: & $x=0$\\
& $x^E = 0$\\
& $r = \Phi$\\
& $f = 1$\\
&\\
Repeat: & $p = A r$\\
& $\alpha = \omega \frac{(p.r)}{(p.p)}$\\
& $f = \frac{f}{1 - \Delta m \alpha} $\\ 
& \\
& $x = x   + \alpha r$\\ 
& $x^E = x^E + f \alpha r$\\
& $r = r   - \alpha p$\\
&\\
\multicolumn{2}{l}{check convergence, continue if necessary}\\
\end{tabular}
\end{center}

\vspace{0.5cm}

Note that the standard overrelaxation parameter $\omega$ can be 
included as usual.

\section{Applications in Lattice QCD}
\label{LAT}

\subsection{The Wilson Fermion Matrix}

In 4 dimensional lattice Quantum Chromodynamics the Wilson fermion 
operator has the structure:
\begin{equation}
M = m - D
\end{equation}
with $m$ being the bare-mass and 
\begin{equation}
D_{x y} = \sum_{\mu=1}^4 (1-\gamma_{\mu}) U_{\mu}(x) \delta_{x,y-\mu}
          + (1+\gamma_{\mu}) U^{\dagger}_{\mu}(x-\mu) \delta_{x,y+\mu}
\end{equation}
the off-diagonal hopping term.  The coefficients ${U_{\mu}(x)}$ are
SU(3) background gauge fields.  Greens Functions in QCD are the
solutions of the linear equation
\begin{equation}
M x = \Phi , \label{greens}
\end{equation}
with $\Phi$ being a $\delta$-source or a 'smeared' source
\cite{SMEARING,FROMMER95}.

\subsection{Results}
\label{RES:NOR}
The numerical tests were performed on the Quadrics Q4 parallel
computer at Wuppertal University and the QH2 at IfH/DESY Zeuthen.  We
test the M$^3$R algorithm on a full-QCD configuration of size $16^3
\times 32$ at $\beta=5.6$ and the following mass trajectory:
\begin{equation}
\kappa = \frac{1}{m} = { 0.1575 , 0.1570 , 0.1565 , 0.1555, 
                         0.1530 , 0.1400 , 0.1000} .
\end{equation}

\begin{figure}[thb]
\epsfbox{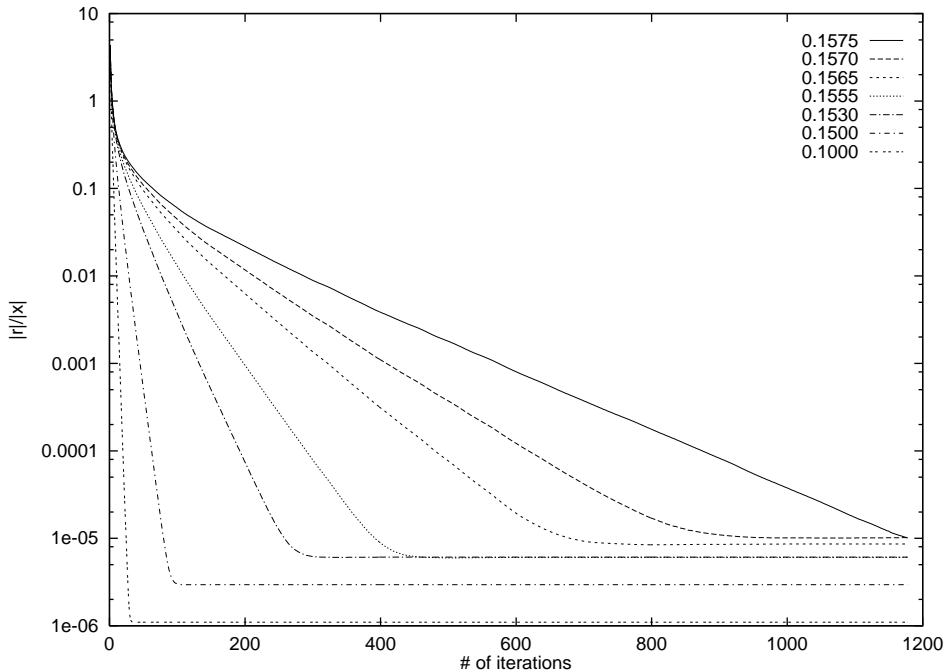}
\caption{True residuals of 7 mass-values against the number of 
         M$^3$R-iterations.\label{normal}}
\end{figure}

We calculated the propagator on the lightest mass, corresponding to
$\kappa=0.1575$ and extrapolated with the M$^3$R - technique to
the six heavier masses. Our stopping criterion was
\begin{equation}
\frac{||r_i||}{||x_i||} < 10^{-5}
\end{equation}
with $||.||$ denoting the 2-norm and $r_i$ being the {\em true}
residual $||M(m+\Delta m) x_i - r_0||$. In figure \ref{normal} the
true residuals of the 7 inversions against the number of M$^3$R
iterations are shown.
It should be noted that the true residuals of the extrapolated heavier
masses are always {\em smaller} than those corresponding to the
lightest mass, for which the MR is actually inverting.  We even
observe a faster convergence with increasing $\Delta m$, despite the
fact that we used a truncated Taylor-expansion to perform the
extrapolation to larger masses.  All residuals reach a plateau
slightly below $10^{-5}$, which reflects the 32bit precision of the
Quadrics computer.

The gain factor for our example on a 256-node QH2 is a factor of 3
in CPU-time compared to seperate inversions at all the masses of
the trajectory. This gain factor, of course, depends on the number of
Green's functions that fit into memory and the mass range of the
trajectory.

\subsection{Even-Odd Preconditioning}
\label{RES:EO}

In lattice quantum chromodynamics it has become standard to use the so
called even-odd or red-black ordering scheme for the lattice-sites.
This ordering scheme induces a block structure of the fermion matrix,
which can be exploited to improve its condition number.

If one denotes the even lattice sites with $x_e$, the odd sites with 
$x_o$, equation \ref{greens} reads
\begin{equation}
  \left( \begin{array}{cc} 
        m       & -D_{eo} \\
        -D_{oe} &   m  
   \end{array} \right) 
  \left( \begin{array}{c}
        x_e \\
        x_o
  \end{array} \right) 
  =    
  \left( \begin{array}{c}
        \Phi_e \\
        \Phi_o
  \end{array} \right)  . 
\end{equation}
Right preconditioning with the matrix
\begin{equation}
M' =
  \left( \begin{array}{cc} 
        m       & D_{eo} \\
        D_{oe} &   m  
   \end{array} \right) 
\end{equation}
yields two seperate matrix equations, each with a smaller condition
number resulting in a performance gain of a factor 2-3:
\begin{equation}
\label{eo_greens}
  \left( \begin{array}{cc} 
        m^2 -D_{eo} D_{oe} & 0 \\
        0 & m^2 -D_{oe} D_{eo}
   \end{array} \right) 
  \left( \begin{array}{c}
        y_e \\
        y_o
  \end{array} \right) 
  =    
  \left( \begin{array}{c}
        \Phi_e \\
        \Phi_o
  \end{array} \right)  .
\end{equation}
The vector $y$ can be transformed back to $x$ using one
additional matrix multiplication:
\begin{equation}
x = M' y . 
\end{equation}

The M$^3$R-Template can be applied as shown above to the even-odd
preconditioned system using $m^2$ as extrapolation parameter.
 
\begin{figure}[thb]
\epsfbox{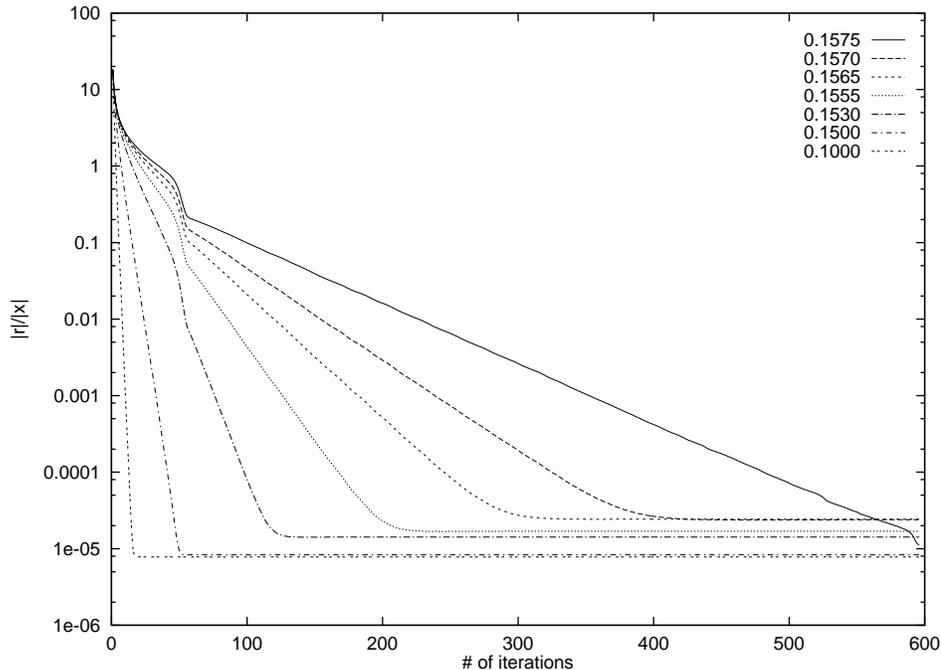}
\caption{True residuals of 7 mass-values against the number
         of M$^3$R-iterations (with even-odd preconditioning).
         \label{evenodd}}
\end{figure}

We test the preconditioned version of the algorithm on the same
configuration, which has been reordered into even-odd form.  We prefer
to use a $\delta$-source on an even lattice-point, because it enables
us to restrict the application to the upper half of the system saving
a factor of two in CPU time. Using the trajectory of seven masses from
section \ref{RES:NOR} we obtain the results shown in fig.
\ref{evenodd}.
The behaviour remains qualitatively the same.  Due to the smaller
condition number of the matrix the convergence speed increases by
approximately a factor of 2.

The gain factor compared to a sequential MR inversion on every mass is
equal to 3. For a general source, however, one has to carry out the
inversion on {\em both} systems of equation \ref{eo_greens} reducing
the above gain factor for our example to 1.5 .

\section{Summary and Outlook}

We have presented a new numerical method to compute a trajectory of
solutions $x$, of the linear system $A x = r$ exploiting the shifted
structure $A = m - D$ of the linear system. We have shown that in a
realistic setup in the framework of lattice Quantum Chromodynamics the
method appears to be numerically stable as well as more efficient than
standard methods. 

For our setup of masses the gain factor varies between 1.5 and 3
depending on the source. Furthermore it depends strongly on the set of
masses to compute. Note that we have chosen a large range of masses to
demonstrate the convergence properties of the algorithm. In a
realistic QCD setup the gain factor would be higher.  This holds in
particular for situations where one would like to make use of the
Feynman-Hellmann theorem with respect to the quark mass dependence as
in the case of the Pion-Nucleon-Sigma term \cite{DONG} in Quenched QCD.

The method can be generalized in many different directions: In
principle it can be embedded into any known iterative solver like
Conjugate Gradient (CG) or Generalized Conjugate Residual (GCR).
Furthermore it seems possible to derive similar algorithms for
matrices with more complex parameter-dependencies than the
one discussed in this paper.

\section{Acknowledgements}
We appreciate the support of the staff at the computer center at
IfH/DESY Zeuthen.

\end{document}